\documentclass{aa}

\usepackage{graphicx}
\usepackage{rotating}
\usepackage{txfonts}
\usepackage{color}
\usepackage{natbib}
\bibpunct{(}{)}{;}{a}{}{,} 

\def\gdor{$\gamma$~Doradus }
\def\Msun{$M_{\odot}$}

\def\Teff{\ensuremath{T_{\mathrm{eff}}}}
\def\cd{d$^{\rm -1}$}
\def\logg{\ensuremath{\log g}}
\def\vmic{$\upsilon_{\mathrm{mic}}$}

\def\vsini{\ensuremath{{\upsilon}\sin i}}
\def\kms{$\mathrm{km\,s}^{-1}$}

\begin{document}

   \title{Constraining the near-core rotation of the $\gamma$ Doradus star 43 Cygni using BRITE-Constellation data\thanks{Based on data collected by the BRITE Constellation satellite mission, designed, built, launched, operated and supported by the Austrian Research Promotion Agency (FFG), the University of Vienna, the Technical University of Graz, the Canadian Space Agency (CSA), the University of Toronto Institute for Aerospace Studies (UTIAS), the Foundation for Polish Science \& Technology (FNiTP MNiSW), and National Science Centre (NCN).}}
   \titlerunning{43 Cygni}

   \author{K. Zwintz \inst{1}\thanks{FWF Elise Richter fellow},
   T. Van Reeth \inst{2,3},
   A. Tkachenko \inst{2},
   S. G\"ossl \inst{1},
   A. Pigulski \inst{4},
   R. Kuschnig \inst{5},
   G. Handler \inst{6}
   A. F. J. Moffat \inst{7}
   A. Popowicz \inst{8},
   G. Wade \inst{9}
          \and
   W. W. Weiss \inst{10}
          }

   \institute{
   Institut f\"ur Astro- und Teilchenphysik, Universit\"at Innsbruck, Technikerstrasse 25/8, A-6020 Innsbruck, Austria\\
              \email{konstanze.zwintz@uibk.ac.at} 
              \and
             Institute of Astronomy, KU Leuven, Celestijnenlaan 200D, B-3001 Leuven
                      \and
                     Kavli Institute for Theoretical Physics, University of California, Santa Barbara, CA 93106, USA 
             \and
Instytut Astronomiczny, Uniwersytet Wroclawski, ul. Kopernika 11, PL-51-622 Wroclaw, Poland
\and 
Institut f\"ur Kommunikationsnetze und Satellitenkommunikation, Technical University Graz, Inffeldgasse 12, A-8010 Graz, Austria 
\and
Nicolaus Copernicus Astronomical Center, ul. Bartycka 18, 00-716, Warsaw, Poland \and 
D\'epartement de physique and Centre de Recherche en Astrophysique du Qu\'ebec (CRAQ), Universit\'e de Montr\'eal, CP 6128, Succ. Centre-Ville, Montr\'eal, Qu\'ebec, H3C 3J7, Canada
\and 
Institute of Automatic Control, Silesian University of Technology, Akademicka 16, 44-100 Gliwice, Poland
\and 
Department of Physics, Royal Military College of Canada, PO Box 17000, Station Forces, Kingston, Ontario, K7K 7B4, Canada
\and 
Institut f\"ur Astrophysik, Universit\"at Wien, T\"urkenschanzstrasse 25/8, A-1180 Vienna, Austria
             }

   \date{Received; accepted}
 
  \abstract
   {Photometric time series of the \gdor star 43 Cyg obtained with the BRITE-Constellation nano-satellites allow us to study its pulsational properties in detail and to constrain its interior structure.}
   {We aim to find a g-mode period spacing pattern that allows us to determine the near-core rotation rate of 43 Cyg and redetermine the star's fundamental atmospheric parameters and chemical composition. }
   {We conducted a frequency analysis using the 156-days long data set obtained with the BRITE-Toronto satellite and employed a suite of MESA/GYRE models to derive the mode identification, asymptotic period spacing and near-core rotation rate. We also used high-resolution, high signal-to-noise ratio spectroscopic data obtained at the 1.2m Mercator telescope with the HERMES spectrograph to redetermine the fundamental atmospheric parameters and chemical composition of 43 Cyg using the software Spectroscopy Made Easy (SME).}
   {We detected 43 intrinsic pulsation frequencies and identified 18 of them to be part of a period spacing pattern consisting of prograde dipole modes with an asymptotic period spacing $\Delta \Pi_{l=1}$ of $2970^{+700}_{-570}\,\rm s$. The near-core rotation rate was determined to be $f_{\rm rot} = 0.56^{+0.12}_{-0.14}\,\rm d^{-1}$. The atmosphere of 43 Cyg shows solar chemical composition at an effective temperature, \Teff, of 7150 $\pm$ 150\,K, a \logg\ of 4.2 $\pm$ 0.6 dex and a projected rotational velocity, \vsini, of 44 $\pm$ 4\,\kms. }
   {The morphology of the observed period spacing patterns shows indications of the presence of a significant chemical gradient in the stellar interior.}


\authorrunning{K. Zwintz et al.}
\titlerunning{Constraining the near-core rotation of 43 Cygni}
   \maketitle

\section{Introduction}
$\gamma$ Doradus stars are intermediate-mass stars, with typical masses between 1.4\,\Msun and 2.0\,\Msun. They cover the transition region between low-mass stars possessing radiative cores and convective envelopes to high-mass stars which have convective cores and radiative envelopes. $\gamma$ Doradus pulsators exhibit non-radial gravity (g) and / or gravito-inertial modes, with observed pulsation periods between 0.3 and 3 days \citep{kaye1999}, that are excited by the convective flux blocking mechanism at the bottom of the convective envelope \citep{guzik2000,dupret2005aa}. These pulsations are most sensitive to the characteristics of the deep stellar interior. The pulsation periods are predicted to be equidistantly spaced in the asymptotic regime ($n\gg l$) for a non-rotating chemically homogeneous star with a convective core and a radiative envelope \citep{tassoul1980}. The presence of a chemical gradient at the edge of the convective core induces non-uniform variations in the spacing pattern \citep{miglio2008}, which typically manifest themselves as periodic dips in the spacings. The periodicity of such non-uniform variations is strongly linked to the location of the chemical gradient inside the star, while their amplitude is related to the steepness of the gradient. 

Stellar rotation leads to shifts in the pulsation frequencies, so that spacing patterns of retrograde modes ($m < 0$) in a rotating star have an overall upward slope, while patterns of zonal and prograde modes ($m > 0$) have a downward slope \citep{bouabid2013}. In addition, rotational mixing, like other extra mixing processes, affects the chemical gradients in the stellar interior, typically reducing the steepness of the gradients, which leads to smaller non-uniform variations in the observed patterns \citep{bouabid2013}. Furthermore, stellar rotation may also affect the propagation cavity of the gravito-inertial modes. While pure g-modes cannot propagate into a convective region, gravito-inertial modes in a rotating star can. This leads to an additional shift of the observed pulsation frequencies \citep{prat2017}.

Thanks to the high-precision photometric observations of space missions such as {\em Kepler} \citep{koch2010}, such period spacing patterns were recently found in some $\gamma$ Doradus stars \citep[e.g.,][]{chapellier2012,kurtz2014,bedding2015,vanreeth2015apjs,keen2015,saio2015}. They have since been used to probe the rotation rates \citep[e.g,][]{murphy2016,vanreeth2016,ouazzani2017,sowicka2017,guo2017} and the nature of the mixing processes \citep[e.g.,][]{schmidaerts2016} in the deep interior of these stars. This provides us with an opportunity to independently evaluate the rotational mixing and angular momentum transport mechanisms inside intermediate-mass stars. Previous studies of the internal rotation profiles using space-based photometric observations have revealed shortcomings in the existing theory for these mechanisms. The observed surface-to-core rotation rates in red giants are orders of magnitude larger than predicted by theory \citep[e.g.,][]{marques2013,cantiello2014}, and \citet{Triana2015} have even found a counter-rotation in the B-type star KIC\,10526294. Internal gravity waves have been proposed as a mechanism to explain these observations \citep{rogers2013,rogers2015}.

43 Cygni (HD 195068/9, V 2121 Cygni, HR 7828) was originally identified as a possible RR Lyrae star from HIPPARCOS photometry \citep{perryman1997}.  \citet{handler1999} indicated the possible \gdor nature of 43 Cyg.
From observed asymmetries in the line profiles, \citet{fekel2003} concluded that 43 Cygni is either a binary or a pulsator. \citet{gerbaldi2007} verified that 43 Cyg is a single star from TD1 UV observations.
\citet{henry2005} identified the star as a (single) $\gamma$ Doradus pulsator and detected three main frequencies at $f_1 = 1.25054$\,\cd, $f_2 = 1.29843$\,\cd\ and $f_3 = 0.96553$\,\cd, using photometric time series in the Johnson $B$ and $V$ filters.
Using 230 spectra obtained over two years, \citet{jankov2006} studied the line profile variations, detected an additional frequency and attempted a first spectroscopic mode identification. The authors reported a frequency at 1.61\,\cd\ to be an $\ell = 5 (\pm 1)$ and $m = 4 (\pm 1)$ and the frequency at 1.25\,\cd\ -- which is $f_1$ of \citet{henry2005} -- to be an $\ell = 4 (\pm 1)$ and $m = 3 (\pm 1)$ mode.
 \citet{cuypers2009} confirmed the presence of the previously identified pulsation frequencies using photometric time series in the Geneva system.
In the literature, 43 Cygni is reported to have an effective temperature, \Teff, of 7300 $\pm$ 250\,K and a \logg\ of 4.35 $\pm$ 0.14\,cms$^{\rm -2}$ \citep{david2015}; its projected rotational velocity, \vsini, was determined to be 44\,\kms \citep[e.g.,][]{fekel2003}.

\begin{figure}[ht]
	\begin{center}
	\includegraphics[width=0.5\textwidth]{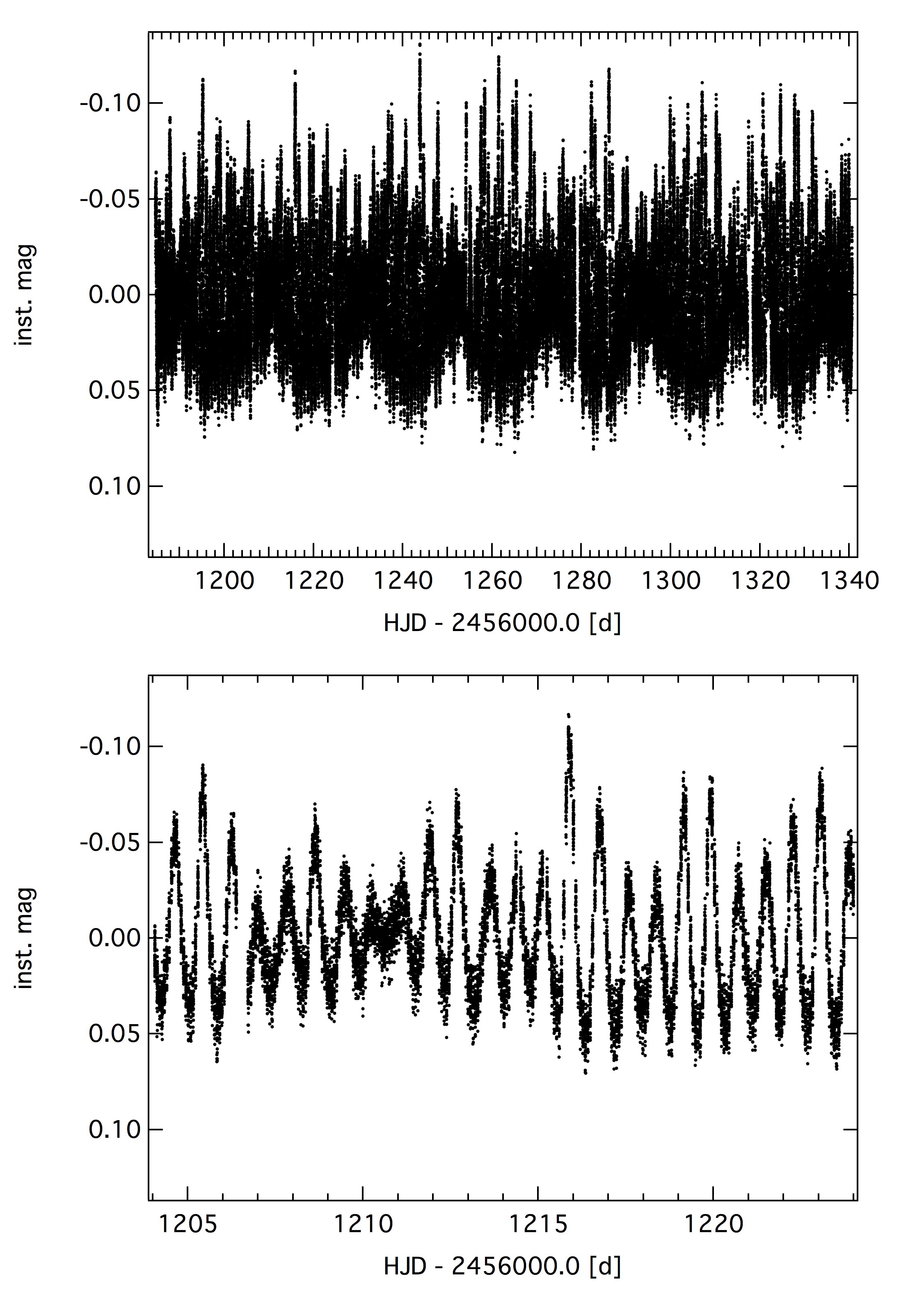}
	\caption{Complete BRITE-Toronto light curve of 43 Cygni (top) and zoom into a 20-day subset (bottom)} 
	\label{fig:lc} 
	\end{center} 
\end{figure}

As 43 Cygni has a $V$ magnitude of 5.74, it is at the fainter end of the magnitude range observable with BRITE-Constellation \citep{weiss2014}. In the present study we describe the BRITE-Constellation observations of 43 Cygni and conduct a detailed asteroseismic investigation of the pulsation periods and their spacings with the aim to constrain the star's near-core rotation. We also use new high-resolution, high signal-to-noise ratio spectroscopy for a re-determination of the fundamental parameters and a first investigation of the chemical abundances of 43 Cygni.

\section{Observations and data reduction}

\subsection{BRITE-Constellation}

BRITE-Constellation\footnote{http://www.brite-constellation.at} is a fleet of five nano-satellites measuring light variations in stars typically brighter than $V=5$ mag. Each 20-cm cube satellite carries a telescope with an aperture of 3\,cm that feeds an uncooled CCD \citep{weiss2014}. Three of the BRITE satellites -- i.e., BRITE-Toronto (BTr), Uni-BRITE (UBr) and BRITE-Heweliusz (BHr) -- carry a custom-defined red filter (550 -- 700\,nm), and two satellites -- i.e., BRITE-Austria (BAb) and BRITE-Lem (BLb) -- carry a custom-defined blue filter (390 -- 460\,nm). More details on the detectors, pre-launch and in-orbit tests are described by \citet{pablo2016}. \citet{popowicz2017} describe the pipeline which processes the observed images yielding the instrumental magnitudes which are delivered to the users. 

BRITE-Constellation observes large fields with typically 15 to 20 stars brighter than $V=6$ mag including at least three targets brighter than $V=3$ mag. Each field is observed at least 15 minutes per each $\sim$100-minute orbit for up to half a year \citep{weiss2014}. 

BRITE-Constellation obtained observations of the Cygnus-II field from June 1 to November 25, 2015. 43 Cygni was observed for only 13 days with BRITE-Lem (BLb) in the blue filter and for 156 days with BRITE-Toronto (BTr) in the red filter, both in chopping mode where the position of the target star within the CCD plane
is constantly alternated between two positions about 20 pixels apart on the CCD. This procedure was adopted to mitigate the impact of high dark current in CCDs. A detailed description of this technique is given in \citet{popowicz2017}. 
As the blue data set from BLb did not have sufficient quality for an asteroseismic analysis, we omitted it from our further investigation.

The BRITE-Constellation data of 43 Cyg have been corrected for instrumental effects according to the iterative procedure described by \citet{pigulski2016}. This procedure was, however, modified to include two-dimensional decorrelations. Since 43 Cyg shows variability with large amplitude, decorrelations were made using residuals from a fit consisting of a sum of sinusoidal terms.
For 43 Cyg the model consisted of five terms: four modes with frequencies 1.250\,\cd, 1.288\,\cd, 0.963\,\cd, and 1.516\,\cd, plus a combination frequency 1.250\,\cd - 0.963\,\cd = 0.287\,\cd). This model was used for all BTr setups. 
The fit was recalculated at each step of the decorrelation. The whole procedure included outlier rejection, one- (1D) and two-dimensional (2D) decorrelations, and removal of corrupted orbits (e.g., affected by poor stability of the satellite). Decorrelations, both 1D and 2D, were made using parameters provided in Data Release 3 \citep[see][or the BRITE Public Data Archive\footnote{https://brite.camk.edu.pl/pub/index.html} for the explanation of parameters]{popowicz2017} and a given satellite orbital phase as an additional parameter. The procedure of decorrelation was made sequentially allowing for multiple decorrelations with the same parameter (or a pair of parameters). At each step, the parameter showing the strongest correlation was chosen for correction. The strength of a correlation was defined as the degree of reduction in variance due to decorrelation with a given parameter (for 1D) or a pair of parameters (for 2D). The iterations were stopped when all correlations (both 1D and 2D) resulted in a variance reduction smaller than 0.05 per cent. The whole procedure was run independently for each setup. The rejection of outliers and the worst orbits was carried out at least twice during the whole procedure. Having decorrelated the data, the blue and red data were separately combined, taking into account the  mean magnitude offsets between the setups.

The complete BTr light curve of 43 Cygni consists of 98481 data points and is shown in the top panel of Fig. \ref{fig:lc} while the bottom panel shows a zoom into a 20-day subset.

\subsection{HERMES spectra}

We obtained ten high-resolution, high signal-to-noise ratio (S/N) spectra for 43 Cygni using the HERMES spectrograph \citep[$R = 85,000$, $\lambda$ = 377 $-$ 900\,nm,][]{raskin2011}  at the 1.2m Mercator telescope (Observatorio del Roque de los Muchachos, La Palma, Canary Islands, Spain). The observations were taken during 7 individual nights between August 4th and 15th, 2016, with typical exposure time of 600\,s and S/N higher than 150 per pixel. 
The individual spectra were reduced with the most recent
version of the HERMES pipeline (release 6) and normalized by fitting a low-order polynomial to carefully selected continuum points. The resulting combined, normalized spectrum has a S/N of 590 calculated from 5820 to 5821\,\AA\ and was used to determine the fundamental parameters and abundances for 43 Cygni. With our spectroscopic data, we also can verify previous reports \citep[e.g.,][]{henry2005,gerbaldi2007} that 43 Cyg is a single star.

\section{Frequency Analysis}

\begin{figure}[t]
	\begin{center}
	\includegraphics[width=0.5\textwidth]{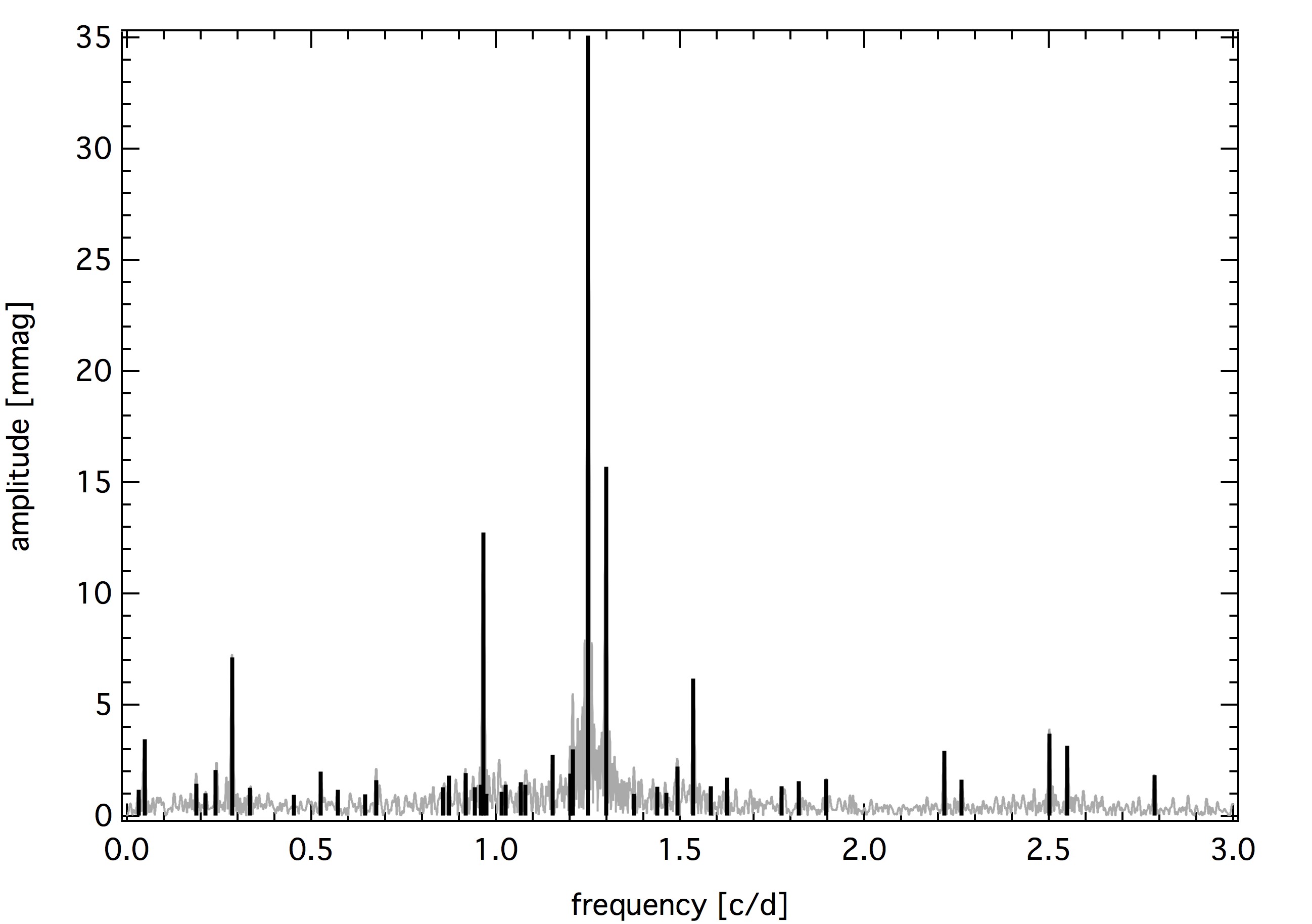}
	\caption{Original amplitude spectrum of 43 Cygni (gray) where the identified pulsation frequencies are marked in black.} 
	\label{fig:ampspec} 
	\end{center} 
\end{figure}

The frequency analysis of the BRITE photometric time series was performed using an iterative prewhitening method based on the Lomb-Scargle periodogram which is described by \citet{vanreeth2015aa}. Frequencies are identified to be significant if they exceed 3.9 times the local noise level in the Fourier domain. 
Frequency errors are calculated using the method by \citet{schwarzenberg2003} which is based on the statistical errors resulting from a non-linear least-squares fit corrected for the correlated nature of the data. Both methods have been already successfully applied to $\gamma$ Doradus pulsation \citep[e.g.,][]{vanreeth2015apjs}.

In total we detected 43 intrinsic frequencies with S/N higher than 3.9 in a range from 0 to 2.78\,\cd\ that are caused by $\gamma$ Doradus type pulsations (see Fig. \ref{fig:ampspec} and Table \ref{tab:freq}). We can also confirm the previously identified frequencies at 1.25054\,\cd, 1.29843\,\cd, 0.96553\,\cd\ and 1.61\,\cd\ reported by Henry et al. (2005) and Jankov et al. (2006).

\begin{table*}
\caption{\label{tab:freq}Pulsation frequencies, periods, amplitudes and signal-to-noise values for 43 Cyg, obtained from the BRITE data, using the iterative prewhitening procedure and sorted by increasing frequency.}
\begin{center}
\begin{tabular}{rrrrrr}
\hline
\hline
\multicolumn{1}{r}{\#} &\multicolumn{1}{c}{frequency} &\multicolumn{1}{c}{period}& \multicolumn{1}{c}{amp} & \multicolumn{1}{c}{S/N} & \multicolumn{1}{c}{comment}   \\
\multicolumn{1}{r}{ } &\multicolumn{1}{c}{[\cd]} &\multicolumn{1}{c}{[d]}& \multicolumn{1}{c}{[mmag]} & \multicolumn{1}{r}{ } & \multicolumn{1}{c}{ }   \\
\hline
1	&	0.03163(16)	&	31.61626(16345)	&	1.192	&	4.69	&	$f_{32}-f_{31}$	\\
2	&	0.04731(8)	&	21.13692(3565)	&	3.449	&	9.07	&	$f_{28}-f_{27}$	\\
3	&	0.18754(14)	&	5.33210(403)	&	1.466	&	4.21	&	$f_{30}-f_{27}$	\\
4	&	0.21185(19)	&	4.72034(423)&	1.013	&	4.04	&	$f_{31}-f_{27}$	\\
5	&	0.24033(11)	&	4.16099(198)&	2.073	&	6.31	&	$f_{32}-f_{27}$	\\
6	&	0.28474(4)	&	3.51203(5)	&	7.146	&	16.56	&	$f_{33}-f_{27} = f_{27}-f_{18}$	\\
7	&	0.33332(16)	&	3.00013(142)&	1.281	&	4.90	&	$f_{34}-f_{27}$	\\
8	&	0.45208(20)	&	2.21202(98)	&	0.947	&	3.94	&	$f_{33}-f_{23}$	\\
9	&	0.52487(12)	&	1.90523(43)	&	1.997	&	6.17	&	$f_{36}-f_{27}$	\\
10	&	0.57092(17)	&	1.75155(51)	&	1.185	&	4.90	&	$f_{37}-f_{27}$	\\
11	&	0.64546(20)	&	1.54928(47)	&	0.974	&	3.90	&	$f_{38}-f_{27}$	\\
12	&	0.67609(14)	&	1.47908(30)	&	1.609	&	5.56	&		\\
13	&	0.85685(15)	&	1.16707(21)	&	1.285	&	4.84	&	$f_{36}-f_{15} = f_{37}-f_{18}$	\\
14	&	0.87263(12)	&	1.14596(16)	&	1.810	&	5.90	&	$f_{38}-f_{21}$	\\
15	&	0.91798(12)	&	1.08935(14)	&	1.933	&	6.02	&		\\
16	&	0.94311(15)	&	1.06032(17)	&	1.289	&	4.70	&		\\
17	&	0.95917(15)	&	1.04257(16)	&	1.418	&	5.56	&		\\
18	&	0.96593(3)	&	1.03527(3)	&	12.758	&	26.60	&	previously identified [{\it 1,2}]	\\
19	&	0.97504(19)	&	1.02559(20)	&	0.994	&	3.91	&		\\
20	&	1.01565(18)	&	0.98459(17)	&	1.086	&	4.20	&	pattern	\\
21	&	1.02591(14)	&	0.97475(14)	&	1.404	&	4.91	&	pattern	\\
22	&	1.06732(14)	&	0.93693(12)	&	1.521	&	5.41	&	pattern	\\
23	&	1.08058(15)	&	0.92543(13)	&	1.407	&	4.93	&	pattern	\\
24	&	1.15385(9)	&	0.86667(7)	&	2.750	&	7.47	&	pattern	\\
25	&	1.20222(12)	&	0.83179(8)	&	1.915	&	5.93	&	$f_{38}-f_{15}$	\\
26	&	1.20847(9)	&	0.82749(6)	&	2.998	&	8.11	&	pattern	\\
27	&	1.25043(2)	&	0.79973(1)	&	35.094	&	48.07	&	pattern; previously identified [{\it 1,2}]  \\
28	&	1.29856(3)	&	0.77008(2)	&	15.699	&	29.54	&	pattern; previously identified [{\it 1,2}]	\\
29	&	1.37508(19)	&	0.72723(10)	&	1.001	&	4.18	&	pattern	\\
30	&	1.43765(15)	&	0.69558(7)	&	1.321	&	4.94	&	pattern	\\
31	&	1.46167(19)	&	0.68415(9)	&	1.036	&	4.18	&	pattern	\\
32	&	1.49255(11)	&	0.66999(5)	&	2.219	&	6.53	&	pattern	\\
33	&	1.53544(5)	&	0.65128(2)	&	6.178	&	15.62	&	pattern	\\
34	&	1.58267(15)	&	0.63184(6)	&	1.340	&	4.87	&	pattern	\\
35	&	1.62701(13)	&	0.61463(5)	&	1.727	&	5.49	&	pattern; previously identified [{\it 1,2}]	\\
36	&	1.77541(15)	&	0.56325(5)	&	1.344	&	5.07	&	pattern	\\
37	&	1.82161(13)	&	0.54896(4)	&	1.570	&	5.38	&	pattern	\\
38	&	1.89567(13)	&	0.52752(4)	&	1.662	&	5.58	&	pattern	\\
39	&	2.21623(9)	&	0.45122(2)	&	2.930	&	8.40	&	$f_{18}+f_{27}$	\\
40	&	2.26298(13)	&	0.44189(3)	&	1.626	&	5.88	&	$f_{18}+f_{28}$	\\
41	&	2.50095(8)	&	0.39985(1)	&	3.717	&	10.24	&	$2f_{27}$	\\
42	&	2.54918(8)	&	0.39228(1)	&	3.157	&	10.06	&	$f_{27}+f_{28}$	\\
43	&	2.78583(12)	&	0.35896(2)	&	1.851	&	7.10	&	$f_{27}+f_{29}$	\\
\hline
\end{tabular}
\end{center}
\tablefoot{References are: {\it 1}: \citet{henry2005}, {\it 2}: \citet{jankov2006}. Pulsations indicated with the comment ``pattern'' are part of the period spacing pattern discussed in Section \ref{sec:asteroseismic}.}
\end{table*}

\section{Spectroscopic analysis}

Spectroscopic analysis of 43 Cyg was performed using the {\sc LLmodels} model atmosphere code \citep{shulyak2004}, the VALD database for atomic line parameters \citep{kupka1999}, SYNTH3 \citep{kochukhov2007} for the computation of synthetic spectra and an updated version of the SME (Spectroscopy Made Easy, version 474) software package \citep{valenti1996,piskunov2017}.

SME allows us to derive effective temperature, surface gravity, overall metallicity, individual element abundances, and the microturbulent, macroturbulent, rotational, and radial velocities of a star by fitting synthetic spectra to those observed. Spectral synthesis calculations may be performed for different grids of model atmospheres and are interpolated between the grid nodes. We used a model grid calculated with the {\sc LLmodels} stellar model atmosphere code for microturbulent velocity \vmic\ = 2.0\,\kms, which ranges from 4500 to 22000\,K in effective temperature, from 
2.5 to 5.0\,dex in surface gravity, and from 
$-$ 0.8 to 0.8\,dex in metallicity \citep[see Table 5 of][]{tkachenko2012}. The corresponding steps are 0.1\,dex in \logg, 0.1\,dex in metallicity, 100\,K in the effective temperature region from 
4500 to 10000\,K and 250\,K for higher \Teff\ values.

For the fitting of the synthetic spectra we followed an approach described for example by \citet{ryabchikova2016} and used the following spectral regions: $4167 - 4511$\,\AA, $4485 - 4590$\,\AA, $4744 - 4983$\,\AA, $5100 - 5200$\,\AA,  $5600 - 5700$\,\AA,  $6100 - 6200$\,\AA, and $6335- 6765$\,\AA. These ranges include the three Balmer lines H$\alpha$, H$\beta$, and H$\gamma$.
For a comparison and consistency check we also used the complete spectral range from $4150 - 6800$\,\AA.

As starting values for our computations, we used the values of \Teff\ = 7300\,K and \logg\ = 4.35 provided by \citet{david2015} and \vsini\ = 44\,\kms\ from \citet{fekel2003}. We used four different approaches to investigate the fundamental parameters for 43 Cyg: (i) we used only the three Balmer lines as input for SME, (ii) we used the four spectral regions without the Balmer lines, (iii) we used all above mentioned regions at the same time and (iv) the complete spectral range from 4\,150--6\,800\AA.
The corresponding solutions are very similar to each other: for \Teff, \logg, and metallicity [Fe/H] we find (i) 7141\,K / 4.11 / $-0.21$, (ii) 7308\,K/ 4.29 / $-0.05$, (iii) 7144 / 4.15 / $-0.18$, and (iv) 7154\,K / 4.24 / $-0.24$. 
All four approaches find \vsini\ values of 44\,$\pm$\,4\,\kms\ and the microturbulent velocity, $v_{\rm mic}$, to vary around a value of 2.9\,\kms. 

\begin{figure}
	\begin{center}
	\includegraphics[width=0.5\textwidth]{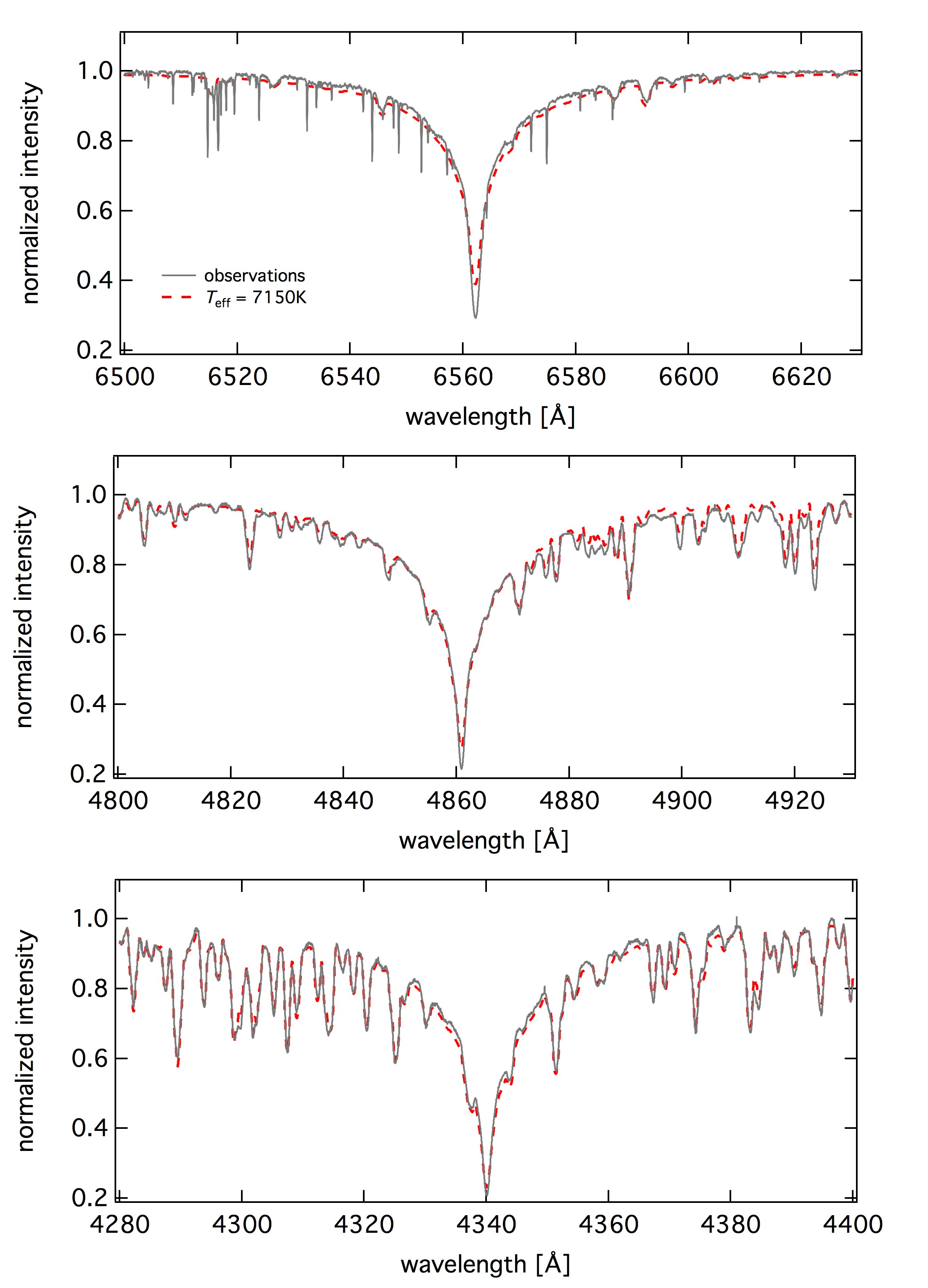}
    \caption{Region of the H$\alpha$ (top), H$\beta$ (middle) and H$\gamma$ (bottom) lines: the observed spectrum is shown in black, and the calculated synthetic spectrum with the final adopted parameters of \Teff = 7150\,K and \logg\ = 4.2 in red.}
    \label{fig:hlines}
	\end{center}
\end{figure}

In the present temperature range, the hydrogen lines are sensitive to changes in both \Teff\ and \logg. We therefore use the average of the results from approaches (i), (iii) and (iv), which include H$\alpha$, H$\beta$, and H$\gamma$ as our final values of \Teff\ and \logg.  
Hence, we adopted the final parameters as \Teff = 7150 $\pm$ 150\,K, \logg = 4.2 $\pm$ 0.6, \vsini = 44 $\pm$ 4\,\kms, and $v_{\rm mic}$ = 2.99 $\pm$ 0.37\,\kms\ (Table \ref{tab:fundpars}). Figure \ref{fig:hlines} shows a comparison between the observed and calculated line profiles of H$\alpha$, H$\beta$ and H$\gamma$.

\begin{table}
\caption{Stellar parameters obtained with SME and estimated values of mass and radius for 43 Cyg}
\label{tab:fundpars}
\begin{center}
\begin{tabular}{ll}
\hline
\hline
\Teff\ & 7150 $\pm$ 150\,K \\
\logg\ & 4.2 $\pm$ 0.6\,dex   \\
\vsini\  &   44 $\pm$ 4\,\kms \\
$[$Fe/H$]$  & -0.05 $\pm$ 0.08\,dex  \\
$v_{\rm micro}$  &  2.99 $\pm$ 0.37\,\kms  \\
\hline
\end{tabular}
\end{center}
\end{table}

\begin{figure}
	\begin{center}
	\includegraphics[width=0.5\textwidth]{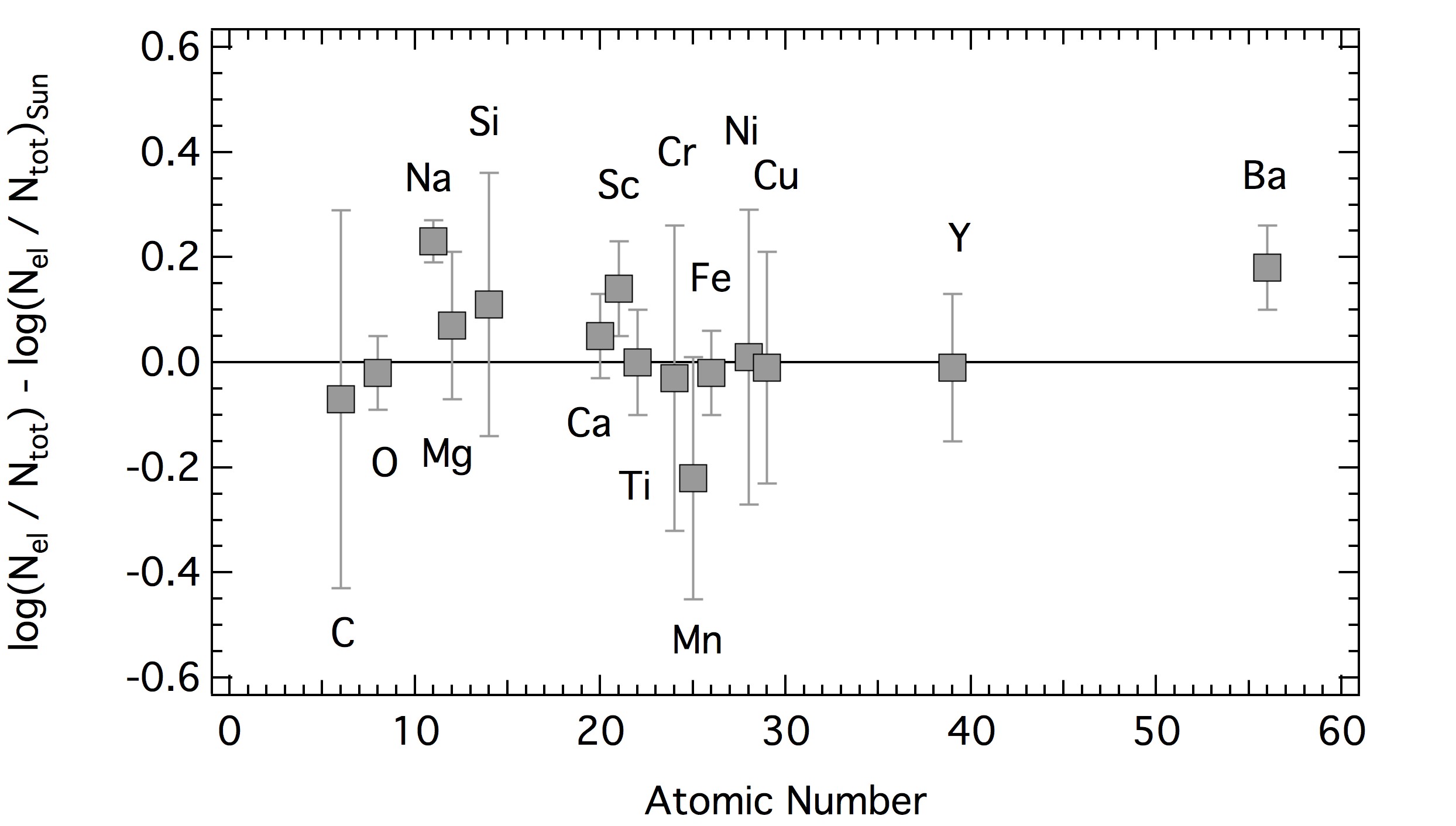}
    \caption{Atmospheric abundances of 43 Cyg relative to the Sun.}
    \label{fig:abund}
	\end{center}
\end{figure}

We also conducted a detailed study of the star's individual chemical abundances using the SME software. We treated each element individually and calculated their atmospheric abundances in LTE except for O, Na, Ca and Ba for which we used a non-LTE analysis based on the procedure of \citet{piskunov2017nlte}. 
Just as we have found for the global metallicity of the star, the individual abundances of all analyzed elements agree with the solar values within the quoted error bars (see Table \ref{tab:abund} and Figure \ref{fig:abund}).
From this analysis we adopt a final value of [Fe/H] of $-0.05$ $\pm$ 0.08 for 43 Cyg.

\begin{table}
\caption{Atmospheric abundances for 43 Cygni with the error estimates based on the internal scattering from the number of the analyzed lines compared to the solar values.}
\label{tab:abund}
\begin{center}
\begin{tabular}{llc}
\hline
\hline
\multicolumn{1}{c}{Element} & \multicolumn{1}{c}{43 Cyg} & \multicolumn{1}{c}{Sun} \\
\multicolumn{1}{c}{ } & \multicolumn{1}{c}{log$(N_{el} / N_{tot}$)} & \multicolumn{1}{c}{log$(N_{el} / N_{tot}$)} \\
\hline
C  & -3.59 $\pm$ 0.36   &  -3.52  \\ 
O & -3.23 $\pm$ 0.07 & -3.21 \\ 
Na & -5.48  $\pm$ 0.04 & -5.71 \\ 
Mg & -4.39 $\pm$ 0.14   & -4.46  \\ 
Si & -4.38  $\pm$ 0.25 & -4.49 \\ 
Ca & -5.63  $\pm$ 0.08 & -5.68 \\ 
Sc &  -8.73 $\pm$ 0.09   &  -8.87 \\ 
Ti & -7.02 $\pm$ 0.10 & -7.02 \\ 
Cr &  -6.40 $\pm$ 0.29  & -6.37 \\ 
Mn &  -6.87 $\pm$ 0.23  & -6.65  \\ 
Fe & -4.56 $\pm$ 0.08 & -4.54 \\ 
Ni & -5.78 $\pm$ 0.28 & -5.79 \\ 
Cu & -7.84 $\pm$ 0.22   & -7.83 \\ 
Y &  -9.81 $\pm$ 0.14   & -9.80  \\ 
Ba &  -9.73 $\pm$ 0.08  & -9.91 \\ 
\hline
\end{tabular}
\end{center}
\tablefoot{Abundances of O, Na, Ca and Ba were calculated in non-LTE based on \citet{piskunov2017nlte}, all other elements in LTE. For comparison the last column gives the abundances of the solar atmosphere calculated by \citet{asplund2009}. }
\end{table}

\section{Asteroseismic inference}
\label{sec:asteroseismic}

\subsection{Identification of a period spacing pattern}
The extracted pulsation frequencies were subsequently converted into periods and analysed to look for period spacing patterns. It is well-known that, because of the typical low noise level, the classical signal-to-noise criterion $S/N\geq4$ is unsuitable for the application of iterative prewhitening on space-based data. 
As for example discussed by \citet{balona2014} and \citet{baran2015}, 
the probability of including spurious extracted frequencies when 
studying space-based data is higher than for ground-based data with 
a lower signal-to-noise ratio. In the case of the latter, the 
classical stop criterion for iterative prewhitening is typically 
fulfilled earlier during the data analysis, so less frequencies are extracted. Spurious extracted frequencies may be 
caused by, e.g., the attempted prewhitening of unresolved pulsation 
frequencies and/or the residuals of imperfect preceding prewhitening. 
Hence, to facilitate the search for period spacing patterns in the data, 
we used the evaluation criterion defined by \citet{vanreeth2015aa} to 
help assess the reliability of the extracted signal, and classified the 
pulsation periods in nine different groups accordingly. In this 
evaluation we compared the amplitudes of the extracted pulsations with 
the local value of the Lomb-Scargle periodogram at that pulsation period. 
The relative difference between these two values was used as a measure 
for the impact of the preceding iterative prewhitening.

\begin{figure*}
	\begin{center}
\includegraphics[width=0.9\textwidth]{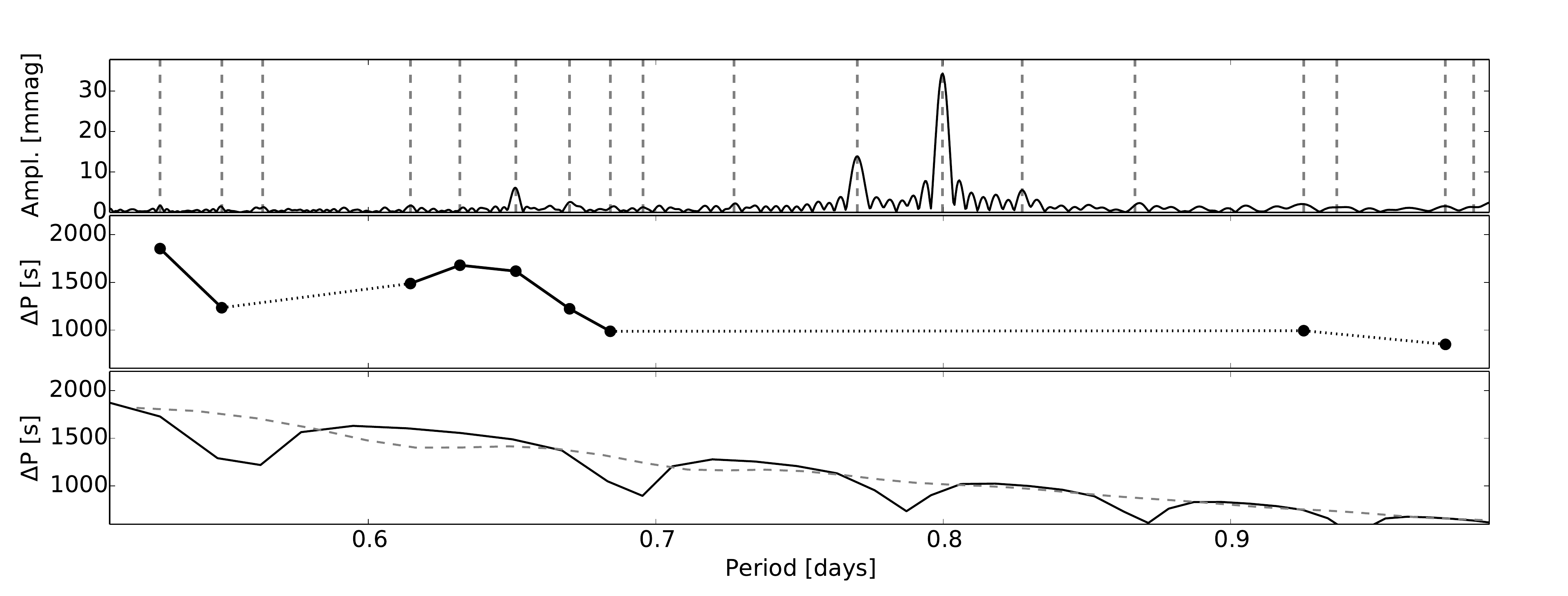}
 	\caption{{\em Top:} the part of the Fourier spectrum where the period spacing pattern was found, with the pulsation periods of the pattern marked by dashed lines. {\em Middle:} the detected period spacing pattern. {\em Bottom:} two model period spacing patterns, for $M = 1.6\,M_\odot$, solar metallicity, $\alpha_{\rm MLT} = 1.8$, $f_{\rm ov} = 0.015\,H_p$ and $X_c = 0.36$. The models have a constant extra diffusive mixing $D_{\rm mix}$ in the radiative region of 0\,$\rm cm^2\,s^{-1}$ (full black) and 1\,$\rm cm^2\,s^{-1}$ (dashed grey), respectively.} 
	\label{fig:43Cyg_spdet} 
	\end{center} 
\end{figure*}

This resulted in the detection of one period spacing pattern, which is shown in Fig.\,\ref{fig:43Cyg_spdet}, and listed in Table\,\ref{tab:freq}. The majority of the extracted pulsation frequencies which are not part of this pattern, were found to be combination frequencies. For the identification of the possible linear combinations we used the frequency resolution of the data set, i.e., $1/T$ = 0.006\cd, as a tolerance margin. This is also listed in Table\,\ref{tab:freq}. In addition, the other detected pulsation frequencies could also be isolated pulsations, possibly with different mode identifications $(l,m)$ from that of the detected pattern, or in the case of some of the lower-amplitude signal, numerical artefacts originating from the iterative prewhitening routine.

\begin{figure*}
	\begin{center}
\includegraphics[width=0.9\textwidth]{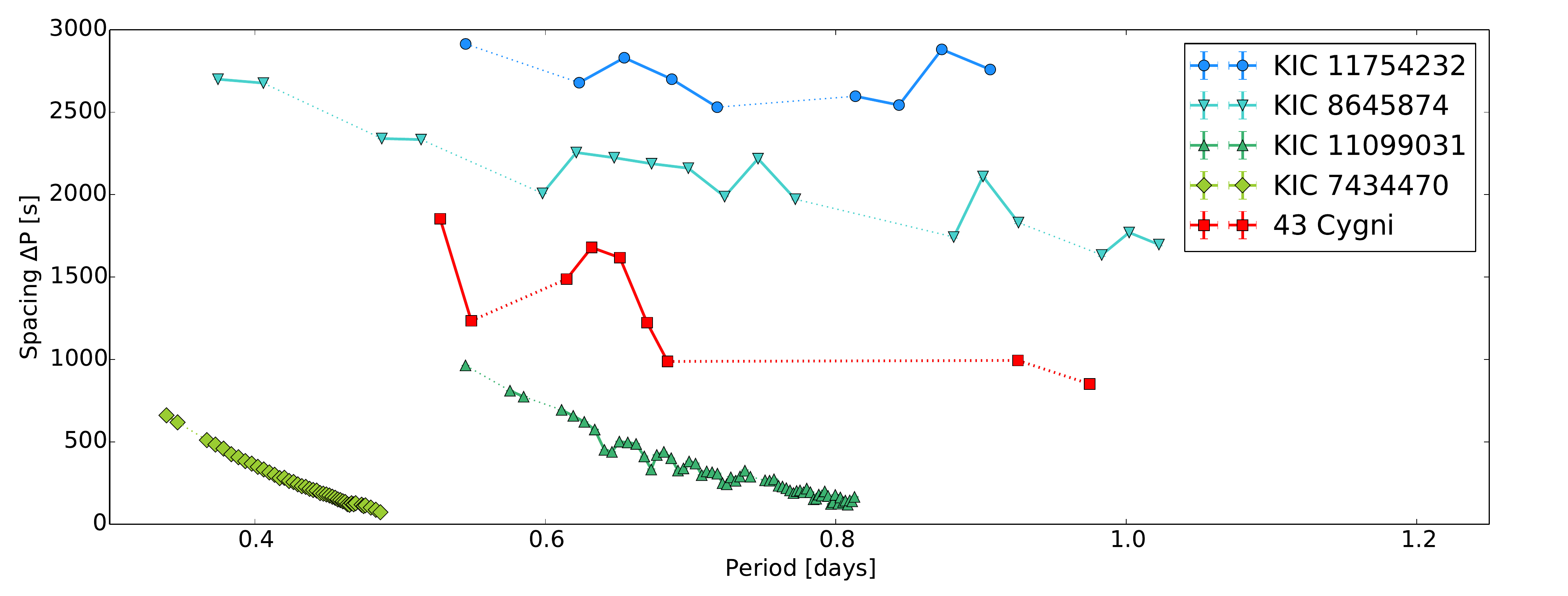}
 	\caption{A comparison of the period spacing pattern of 43\,Cyg with a selection of period spacing patterns detected for $\gamma$\,Dor stars observed with the {\em Kepler} space mission. THe non-uniform variations of the period spacings of 43\,Cyg, caused by chemical gradients in the stellar interior, are clearly visible. The tilt in the pattern, caused by the stellar rotation, follows the trend established by the other observations.} 
	\label{fig:compar} 
	\end{center} 
\end{figure*}

Although the detected pattern is relatively short and has gaps, there is a clear downward slope, caused by the stellar rotation. The pattern also contains strong non-uniform variations, indicative of a chemical gradient in the deep stellar interior. This is especially obvious when the pattern is compared to previously detected patterns, as shown in Figure\,\ref{fig:compar}. 
While a detailed asteroseismic modelling would require a more 
extensive period spacing pattern, and thus, a significantly longer 
time base of observations, the current data allow for a basic 
asteroseismic evaluation. We can \emph{(i)} determine the mode 
identification $(l,m)$ of the pulsation modes in the period spacing 
pattern, and compute the asymptotic period spacing
 $\Delta\Pi_{l=1}$ and near-core rotation rate $f_{\rm rot}$ from the observed period spacing pattern (see Sect. \ref{mode-id}); and {\em (ii)} evaluate the consistency between the obtained spectroscopic properties of the star and the derived asymptotic period spacing and near-core rotation rate, using a suite of MESA/GYRE models (see Sect. \ref{comparison}).

\begin{figure*}[ht]
	\begin{center}
	\includegraphics[width=\textwidth]{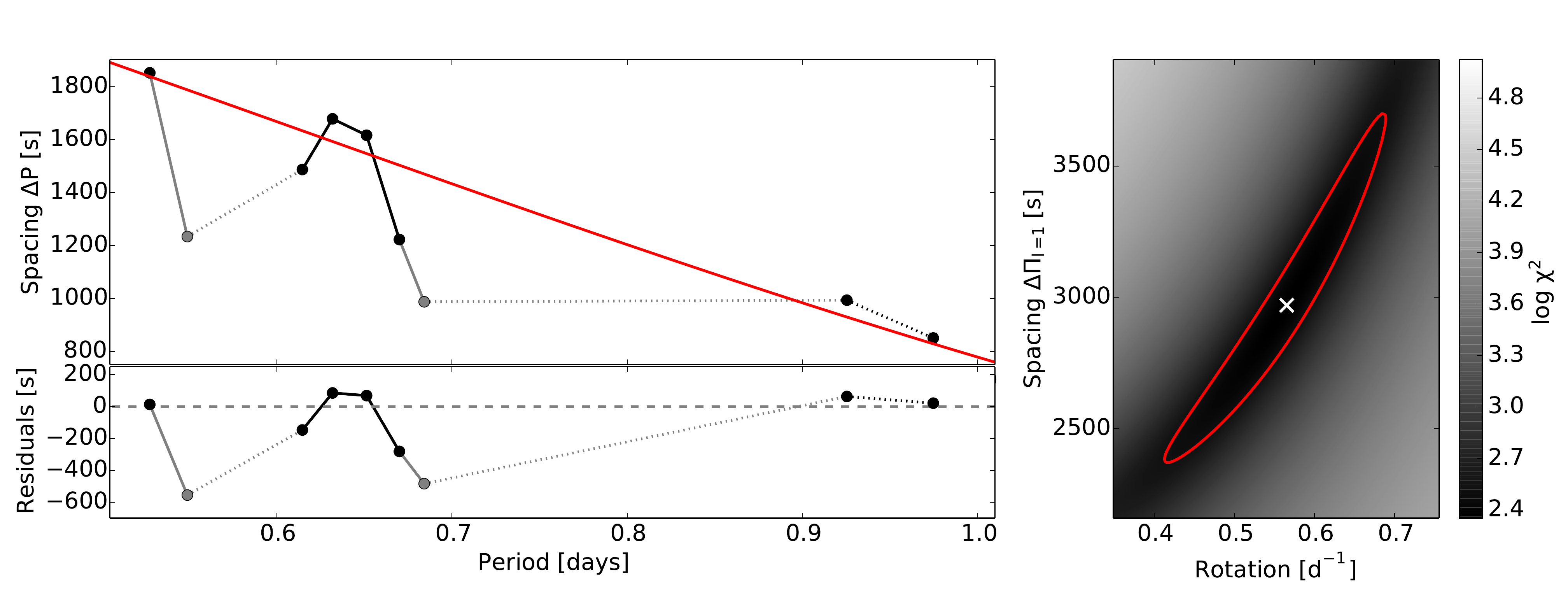}
	\caption{{\em Top left:} the observed period spacing pattern (black) with the best asymptotic model pattern (full red line). The largest dips in the pattern (gray) were not taken into account for the model fit. {\em Bottom left:} the residuals of the fit of the model to the observed period spacings. {\em Right:} the $\chi^2$-landscape of the fit, with the location of the best model (white cross) and the $1-\sigma$ confidence region (red contour).}
	\label{fig:43Cyg_rotfit} 
	\end{center} 
\end{figure*}

\begin{figure*}[ht]
	\begin{center}
	\includegraphics[width=\textwidth]{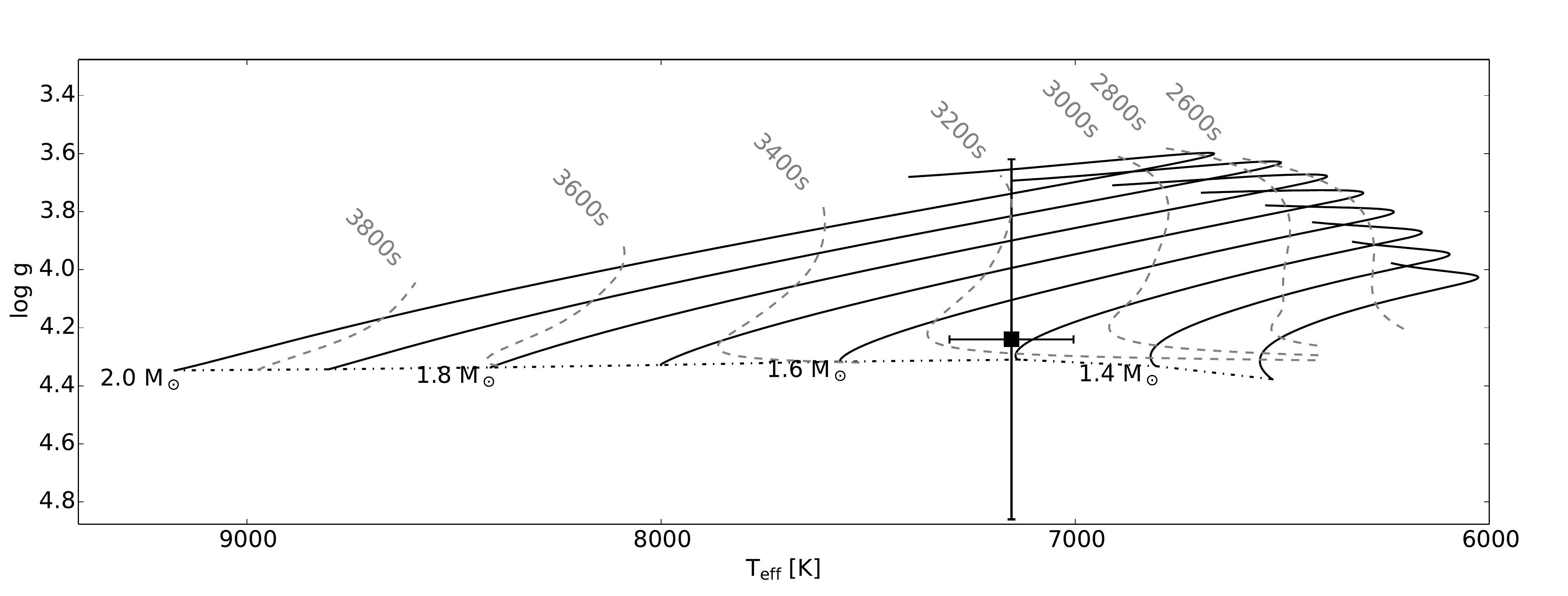}
	\caption{The location of 43\,Cyg (black square) in a Kiel diagram. The evolution tracks (full black lines) have been computed with MESA v7385, assuming solar metallicity, and exponential overshooting $f_{ov} = 0.015$. Both the ZAMS (black dash-dotted line) and the values of the asymptotic spacing $\Delta\Pi_{l=1}$ (gray dashed lines) are also shown.} 
	\label{fig:kiel} 
	\end{center} 
\end{figure*}

\subsection{Mode-identification and near-core rotation rate}
\label{mode-id}
In the first part of our analysis, we subsequently fitted a model 
spacing pattern to these observations to yield a mode identification 
for these pulsations and derive the near-core rotation rate. In this 
model, we ignored the non-uniform variations in period spacing patterns 
caused by chemical gradients, and assumed a rigidly rotating chemically 
homogeneous star. We used the asymptotic equidistant period spacing pattern, 
derived by \citet{tassoul1980}, and included the effects of rotation using 
the traditional approximation \citep[e.g.][]{eckart1960,lee1987,lee1997}. 
The fit of the model pattern to the observations was evaluated using a 
$\chi^2$-statistic. We refer the reader to \citet{vanreeth2016} for a 
detailed description and validation of this method.

For the evaluation of this basic model pattern, we excluded the pulsation periods which are strongly affected by the chemical gradients in the stellar interior, at 0.55\,d and 0.68\,d.  The presence of a chemical gradient inside the star leads to the trapping of pulsation modes \citep{miglio2008}, and as we can clearly see in Figure\,\ref{fig:43Cyg_spdet}, these two pulsation periods are responsible for the presence of the dips in the detected period spacing pattern. The traditional approximation was implemented using the TAR module of the one-dimensional stellar pulsation code GYRE v5.0 \citep{townsend2013}. We identified the pulsations as prograde dipole modes, and found a near-core rotation frequency $f_{\rm rot} = 0.56^{+0.12}_{-0.14}\,\rm d^{-1}$ and an asymptotic spacing $\Delta \Pi_{l=1} = 2970^{+700}_{-570}\,\rm s$. This is in full agreement with the results of previous studies of $\gamma$\,Dor stars in the literature \citep[e.g.,][]{vanreeth2016,ouazzani2017}. This result is also shown in Fig.\,\ref{fig:43Cyg_rotfit}.

\subsection{Comparison between spectroscopic and asteroseismic properties}
\label{comparison}
In the second part of this analysis, we made a consistency check between the spectroscopic and photometric parameters, i.e., $T_{\rm eff}$, $\log\,g$, $[M/H]$, $\Delta\Pi_{l=1}$ and $f_{\rm rot}$, using stellar models computed with the stellar evolution code MESA v7385 \citep{paxton2011,paxton2013,paxton2015}. We found that, in part because of the large error margins on the different parameters, the derived parameter values are entirely consistent. This is illustrated in the Kiel diagram in Fig.\,\ref{fig:kiel}. The stellar evolution tracks in this Figure were computed assuming solar metallicity, using the solar abundances derived by \citet{asplund2009}. The convection was treated using the mixing-length theory, with $\alpha_{\rm MLT}$ equal to 1.8, and the models include exponential core overshooting, assuming $f_{ov}$ equal to 0.015. These parameter values were chosen to be close to typical values found in the literature for main-sequence g-mode pulsators \citep[e.g.,][]{moravveji2015,moravveji2016,schmidaerts2016}.

\subsection{Morphology of the period spacing pattern and diffusive mixing}
In addition, the morphology of the observed pattern also has some interesting characteristics. The large dips in the observed period spacing pattern point to the presence of a significant chemical gradient in the stellar interior. The relative amplitude of the dips in this pattern is comparable to those of the largest dips previously observed in period spacing patterns of $\gamma$\,Dor stars \citep[e.g.,][]{bedding2015,vanreeth2015apjs}. This is a possible indication that the amount of chemical mixing in this star is on the lower end of what can be expected for a typical $\gamma$\,Dor star.

Hence, we have made an additional consistency check, whereby we qualitatively compare the observed data with two theoretical patterns, that were calculated with the stellar pulsation code GYRE, using stellar models computed with the stellar evolution code MESA as input. This is illustrated in the bottom panel of Fig.\,\ref{fig:43Cyg_spdet}. Both models were selected in agreement with the values of $T_{\rm eff}$, $\logg$ and $\Delta \Pi_{l=1}$ within the $1\sigma$ confidence intervals. They have a stellar mass of 1.6\,$M_\odot$, 
solar metallicity, a convective core exponential overshoot of 0.015 and a central hydrogen fraction $X_c$ of 0.36. The two models have a constant extra diffusive mixing $D_{\rm mix}$ in the radiative region of 0\,$\rm cm^2\,s^{-1}$ and 1\,$\rm cm^2\,s^{-1}$, respectively. In these models, the extra diffusive mixing reduces the steepness of the chemical gradients in the stellar interior. It seems that for these models, an extra mixing $D_{\rm mix}$ of $1\,\rm cm^2\,s^{-1}$ is an upper limit for 43\,Cyg. Thus, we speculate that one or more steep chemical gradients are present in the interior of 43\,Cyg. Detailed asteroseismic modelling is required to accurately determine the amount of extra mixing present inside the star, and evaluate the validity of these models.  However, this is outside of the scope of this paper.

\section{Summary and conclusions}

We have identified a period spacing pattern for the \gdor star 43 Cyg from the 156-days long photometric time series obtained by the BRITE-Toronto nano-satellite which can be used to constrain the star's near-core rotation. We also used high-resolution spectroscopic data to redetermine the star's fundamental parameters to be \Teff\ = 7150 $\pm$ 150\,K, \logg\ = 4.2 $\pm$ 0.6, \vsini\ = 44 $\pm$ 4\,\kms, and $v_{\rm micro}$ = 2.99 $\pm$ 0.37\,\kms. 43 Cyg's atmosphere shows solar chemical composition with no peculiarities.

Of the 43 identified pulsation frequencies in 43 Cyg, 18 are part of the identified period spacing pattern, while 19 frequencies could be explained as combination frequencies. The remaining pulsation frequencies are possible isolated pulsations with different mode properties than those of the detected pattern or are part of a different, yet unidentified spacing pattern.

43 Cyg's detected period spacing pattern clearly shows a downward slope and strong non-uniform variations which indicate a chemical gradient in the stellar interior. We identified the 18 modes of the pattern to be prograde dipole modes and used them to find 43 Cyg's near-core rotation rate to be $f_{\rm rot} = 0.56^{+0.12}_{-0.14}\,\rm d^{-1}$ and its asymptotic period spacing $\Delta \Pi_{l=1}$ to amount to $2970^{+700}_{-570}\,\rm s$. 
The relative amplitude of the dips in the observed period spacing pattern is indicative of one or more strong chemical gradients in the interior of 43\,Cyg. Indeed, theoretical patterns computed with MESA/GYRE for the cases without extra diffusive mixing and with low amount of mixing $D_{\rm mix}$ of $1\,\rm cm^2\,s^{-1}$, which reduces the steepness of the chemical gradients in the models, show a significant difference in the shape of the dips that range from very pronounced to strongly washed out.

The limits of our analysis are determined by the relatively short detected period spacing pattern that also shows gaps. A detailed asteroseismic modeling can only be conducted based on a longer pattern derived from a significantly longer time base of observations.

\begin{acknowledgements}
KZ acknowledges support by the Austrian Fonds zur F\"orderung der wissenschaftlichen Forschung (FWF, project V431-NBL).  TVR gratefully acknowledges support from the Fund for Scientific Research of Flanders (FWO), Belgium, under grant agreement G.0B69.13. SG was supported by the University of Innsbruck through the "Nachwuchsf\"orderung 2015" project entitled "Asteroseismology with BRITE-Constellation'' (PI: K. Zwintz). The research leading to these results has (partially) received funding from the European Research Council (ERC) under the European Union’s Horizon 2020 research and innovation programme (grant agreement N$^\circ$670519: MAMSIE). TVR is grateful to the Kavli Institute of Theoretical Physics at the University of California, Santa Barbara, for the kind hospitality during the scientific research visit, during which part of the present research was conducted, and which was supported by the Fund for Scientific Research of Flanders (FWO), Belgium (grant agreement V427217N) and in part by the National Science Foundation under Grant No. NSF PHY11-25915. We are grateful to Bill Paxton and Richard Townsend, and their collaborators at the Universities of California at Santa Barbara and Wisconsin-Madison, for their valuable work on the stellar evolution code MESA and stellar pulsation code GYRE.
A. Pigulski and A. Popowicz acknowledge support from the Polish National Science Centre (grant No. 2016/21/B/ST9/01126 and 2016/21/D/ST9/00656). The Polish contribution to the BRITE mission is funded by the Polish National Science Centre (NCN, grant 2011/01/M/ST9/05914) and a SPUB grant by the Polish Ministry of Science and Higher Education. GH acknowledges support by the Polish NCN grant 2015/18/A/ST9/00578.
GAW acknowledges Discovery Grant support from the Natural Sciences and Engineering Research Council (NSERC) of Canada. A.F.J.M. is grateful for financial aid from NSERC (Canada) and FQRNT (Quebec). 
\end{acknowledgements}


\bibliographystyle{aa}
\bibliography{43Cyg.bib}

\end{document}